\documentclass[twocolumn,showkeys,byrevtex,superscriptaddress,prl,nofootinbib]{revtex4}%
\pdfoutput=1
\usepackage{amsfonts}
\usepackage{amsmath}
\usepackage{amssymb}
\usepackage{tikz}
\usetikzlibrary{calc}
\usetikzlibrary{arrows}
\usepackage{tikz-3dplot}
\usepackage{graphicx}
\usepackage{times}

\newcommand*{\ArcAngle}{60}%
\newcommand*{\ArcRadius}{2.0}%

\newtheorem{theorem}{Theorem}

 \newcommand{\ket}[1]{|#1\rangle}

\begin{document}
\preprint{ }
\title{Summoning Information in Spacetime, or \\
Where and When Can a Qubit Be?}
\author{Patrick Hayden}
\affiliation{\textit{Stanford Institute for Theoretical Physics, Physics Department, Stanford University, CA 94304-4060, USA }}
\affiliation{\textit{Perimeter Institute for Theoretical Physics, Waterloo, Ontario, Canada}}
\author{Alex May}
\affiliation{\textit{Department of Physics and Astronomy, University of British Columbia, Vancouver, BC, V6T 1W9, Canada}}

\begin{abstract}
One of the most important properties of quantum information, and the one ultimately responsible for its cryptographic applications, is that it can't be copied. That statement, however, is not completely accurate.
While the no-cloning theorem of quantum mechanics prevents quantum information from being copied in space, the reversibility of microscopic physics actually \emph{requires} that the information be copied in time.
In spacetime as a whole, therefore, quantum information is widely replicated but in a restricted fashion. We fully characterize which regions of spacetime can all hold the same quantum information.
Because quantum information can be delocalized through quantum error correction and teleportation, it need not follow well-defined trajectories. 
Instead, replication of the information in any configuration of spacetime regions not leading to violations of causality or the no-cloning principle is allowed.
To demonstrate this, we answer the operational question of exactly when the information can be summoned to a set of spacetime points, showing how to do so efficiently using a combination of teleportation and codeword-stabilized quantum codes.
This provides a simple and complete description of where and when a qubit can be located in spacetime, revealing a remarkable variety of possibilities.
\end{abstract}

\date{16 February 2016}
\startpage{1}
\endpage{10}
\maketitle

To understand how information can be distributed through spacetime, it is important to understand the trajectories of any possible physical information carriers. For example, a bit of classical information encoded into a beam of light at a particular point in spacetime can travel anywhere the light can. 
The set  of possible spacetime destinations of that bit is therefore known as the point's future light cone. By copying the bit as necessary, the future light cone can essentially be filled with copies of the bit, subject only to fundamental limits on spacetime information density like the holographic bound~\cite{bousso02}.

Quantum information is very different. While a qubit's possible destinations are still restricted to lie within its future light cone, the no-cloning theorem prevents the qubit from being copied~\cite{wootters1982single}. This suggests that a qubit must simply follow a single physically reasonable trajectory within the light cone, which would indeed be true if the qubit were required to be localized. However, there is no such requirement. A qubit can be stored in the long-range correlations of condensed matter systems~\cite{kitaev2003}, delocalized using quantum error correction~\cite{gottesman97} and even teleported~\cite{bennett92}. The purpose of this article will therefore be to understand how a quantum state can be delocalized and subsequently \emph{re}localized, and thereby exactly how quantum information can be distributed through spacetime. 

If some quantum information is localized in the vicinity of a spacetime point, it should be possible in principle to quickly assemble and exhibit that information nearby. Indeed, that is the very {definition} of what it means for the information to be localized. To track propagating information, it then simply suffices to identify the extent to which the information is localized about a collection of points in spacetime. That idea can be rigorously formalized using a version of the information processing task called \emph{summoning}, whose importance for relativistic quantum theory has already been identified by Kent~\cite{kent11,kent12}.
Suppose that some quantum information, in the form of a quantum system in an unknown quantum state $\ket{\varphi}$, is initially localized at some spacetime point $s$. A request for the state will be received at one of a set of possible spacetime \emph{call points} $\{ y_0, y_1, \ldots, y_{n-1} \}$.
Associated to each call point $y_j$ is a \emph{reveal point} $z_j$ in its future light cone, with the rule that if the state is requested at $y_j$ then it must be revealed at $z_j$. The more separated $y_j$ and $z_j$ are, the more delocalized the information can be between $y_j$ and $z_j$. However, much like performing a quantum measurement, exhibiting the information at a specific $z_j$ will eliminate all indeterminacy about its position. Therefore, it is sufficient to be able to respond to the request \emph{in principle}, which can be enforced by stipulating that at most one of the $n$ call points will receive the request, but which one isn't known in advance.\footnote{A more extended discussion of the assumptions implicit in this definition can be found in Appendix A.}
See Figure~\ref{fig:simple} for an illustration.
\begin{figure}
\begin{tikzpicture}[scale=0.7]

\coordinate (P) at (0cm, 0cm);

\coordinate (C2) at (2.5cm, 2.5cm);
\coordinate (Q2) at (2.5cm, 4.5cm);
\coordinate (L2) at (3.5cm,3.5cm);
\coordinate (R2) at (1.5cm,3.5cm);

\coordinate (C1) at (-2.5cm,2.5cm);
\coordinate (Q1) at (-2.5cm,4.5cm);
\coordinate (L1) at (-3.5cm,3.5cm);
\coordinate (R1) at (-1.5cm,3.5cm);

\draw[thick] (C1) -- (Q1); 	 
\draw[thick] (C2) -- (Q2); 

\draw[fill=yellow] (P) circle (0.15cm);
\node[below right] at (P) {$s$};

\draw[fill=black] (C1) circle (0.15cm);
\node [below right] at (C1) {$y_0$};
\draw[fill=black] (C2) circle (0.15cm);
\node [below right] at (C2) {$y_1$};
	
\draw[fill=blue] (Q1) circle (0.15cm);
\node [above right] at (Q1) {$z_0$};
\draw[fill=blue] (Q2) circle (0.15cm);
\node [above right] at (Q2) {$z_1$};

\draw[->] (-3.5,0) -> (-3.5,1);
\draw[->] (-3.5,0) -> (-2.5,0);

\node [left] at (-3.5,1) {t};
\node [below] at (-2.5,0) {x};

\end{tikzpicture}
\caption{ Spacetime diagram of a summoning task with call and reveal points separated. The state is initially localized at $s$ and will be called at one of $y_0$ or $y_1$, in which case the state must be revealed at $z_0$ or $z_1$, respectively.}\label{fig:simple}
\end{figure}
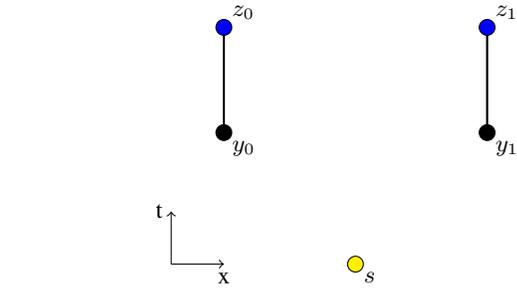

In characterizing the relationship between localization and summoning, it is actually possible to be more precise. Define the \emph{causal diamond} $D_j$ to be the intersection of the future light cone of $y_j$ and the past light cone of $z_j$.\footnote{The usual definition of a causal diamond would require that $z_j$ be timelike to $y_j$ but we relax that requirement here.} This represents the region in which it is possible to both be affected by a request at $y_j$ and affect the outcome at $z_j$. A qubit can be summoned to $z_j$ from $y_j$ if and only if the qubit is localized in the causal diamond $D_j$.\footnote{When $z_j$ is timelike to $y_j$, the causal diamond $D_j$ can be foliated by spacelike hypersurfaces related by unitary time evolution. If the summoning request can be satisfied, the quantum information must therefore be available in the degrees of freedom entering $D_j$ through the future light cone of $y_j$.} Therefore, the summoning task is possible if and only if the qubit's information is replicated in each and every one of the causal diamonds $\{ D_1, D_2, \ldots, D_{n-1} \}$. (For linguistic convenience, we always discuss summoning a qubit, but our results extend trivially to any finite-dimensional quantum information.)

To understand the definition and the surprising variety phenomena it allows, it will be helpful to consider some important examples.
The simplest one consists of just two call points coinciding precisely with their associated reveal points, and in the future light cone of a qubit in the state $\ket{\varphi}$ at the starting point. If the two call points are not spacelike to each other, then the  qubit can simply be transmitted to each call point in turn. On the other hand, Kent observed that if the two call points are spacelike to each other, the impossibility of superluminal signalling implies that being able to successfully summon $\ket{\varphi}$ would amount to being able to send $\ket{\varphi}$ to both the reveal points, which is a clear violation of the no-cloning principle. Despite its simplicity, this no-summoning theorem has significant consequences for information processing. In the same way that the no-cloning theorem gives rise to secret key distribution protocols secured by the laws of quantum mechanics, the no-summoning theorem gives rise to secure bit commitment protocols secured by a combination of quantum mechanics and relativity~\cite{kent2011b}. Bit commitment, a cryptographic primitive which can be used to build secure distributed computations, is impossible using quantum mechanics alone~\cite{mayers97,lo97}.

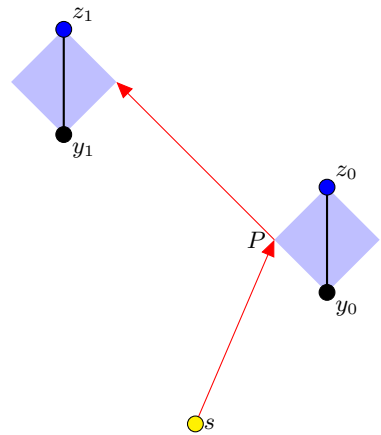
\begin{figure}[t]
\begin{tikzpicture}[scale=0.7]

\coordinate (P) at (0cm, 0cm);

\coordinate (C1) at (2.5cm, 2.5cm);
\coordinate (Q1) at (2.5cm, 4.5cm);
\coordinate (L1) at (3.5cm,3.5cm);
\coordinate (R1) at (1.5cm,3.5cm);

\coordinate (C2) at (-2.5cm,5.5cm);
\coordinate (Q2) at (-2.5cm,7.5cm);
\coordinate (L2) at (-3.5cm,6.5cm);
\coordinate (R2) at (-1.5cm,6.5cm);

\foreach \i in {1,2}
	{
	\fill[blue!25] (C\i) -- (L\i) -- (Q\i) -- (R\i);
	}

\draw[thick] (C1) -- (Q1); 	 
\draw[thick] (C2) -- (Q2); 

\draw[fill=black] (C1) circle (0.15cm);
\node [below right] at (C1) {$y_0$};
\draw[fill=black] (C2) circle (0.15cm);
\node [below right] at (C2) {$y_1$};
	
\draw[fill=blue] (Q1) circle (0.15cm);
\node [above right] at (Q1) {$z_0$};
\draw[fill=blue] (Q2) circle (0.15cm);
\node [above right] at (Q2) {$z_1$};

\node[left] at (R1) {$P$};

\draw[red][-triangle 45] (P) -- (R1);
\draw[red][-triangle 45] (R1) -- (R2);

\draw[fill=yellow] (P) circle (0.15cm);
\node[right] at (P) {$s$};

\end{tikzpicture}
\caption{Transmission through causal diamonds.
  In this example, a simple strategy will work even though $z_1$ is not in the future light cone of $z_0$. The quantum state is first transported along the arrow to $P$. The call information originating at $y_0$ is broadcast into its future light cone and accessed at the point $P$. If the call is for $z_0$, the quantum state is moved there. If not, it is moved to $z_1$.}\label{fig:diamond}
\end{figure}

Another simple example in $1+1$ dimensions is shown in Figure~\ref{fig:diamond}. Even though there is no causal curve through each of the reveal points, there is a causal curve passing through each of the causal diamonds, which is sufficient to complete the summoning task.

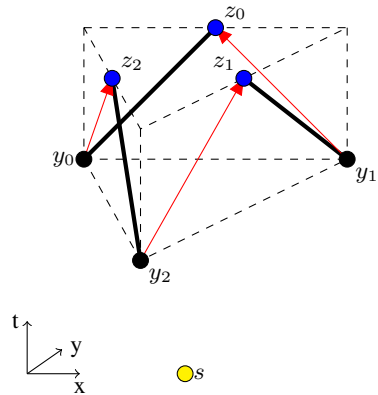
\begin{figure}[t]
\begin{tikzpicture}[scale=0.7]

	\coordinate (PS) at (2.5,-3.5,1.5);

	\coordinate (Ca) at (0,0,0);
	\coordinate (Cb) at (5,0,0);
	\coordinate (Cc) at (3,0,5);
	
	\coordinate (Qa) at (2.5,2.5,0);
	\coordinate (Qb) at (4, 2.5, 2.5);
	\coordinate (Qc) at (1.5, 2.5, 2.5);
	
	\draw[dashed] (Ca) -- (Cb) -- (Cc) -- (Ca);
	\draw[dashed] (0,2.5,0) -- (5,2.5,0) -- (3,2.5,5) -- (0,2.5,0);
	\draw[dashed] (0,2.5,0) -- (0,0,0);
	\draw[dashed] (5,2.5,0) -- (5,0,0);
	\draw[dashed] (3,2.5,5) -- (3,0,5);
	
	\draw[ultra thick] (Ca) -- (Qa);
	\draw[ultra thick] (Cb) -- (Qb);
	\draw[ultra thick] (Cc) -- (Qc);
	
	\draw[-triangle 45][red] (Ca) -- (Qc);
	\draw[-triangle 45][red] (Cb) -- (Qa);
	\draw[-triangle 45][red] (Cc) -- (Qb);
	
	\draw[fill=blue] (Qa) circle (0.15);
	\draw[fill=blue]  (Qb) circle (0.15);
	\draw[fill=blue]  (Qc) circle (0.15);
	
	\node [above right] at (Qa) {$z_0$};
	\node [above left] at (Qb) {$z_1$};
	\node [above right] at (Qc) {$z_2$};
	
	\draw[fill=black] (Ca) circle (0.15);
	\draw[fill=black] (Cb) circle (0.15);
	\draw[fill=black] (Cc) circle (0.15);
	
	\node [left] at (Ca) {$y_0$};
	\node [below right] at (Cb) {$y_1$};
	\node [below right] at (Cc) {$y_2$};
	
	\draw[fill=yellow] (PS) circle (0.15);
	\node [right] at (PS) {$s$};
	
	\draw[->] (-0.5,-3.5,1.5) -> (0.5,-3.5,1.5);
	\draw[->] (-0.5,-3.5,1.5) -> (-0.3,-3.5,0.3);
	\draw[->] (-0.5,-3.5,1.5) -> (-0.5,-2.5,1.5);
	
	\node[below] at (0.5,-3.5,1.5) {x};
	\node[right] at (-0.3,-3.5,0.3) {y};
	\node[left] at (-0.5,-2.5,1.5) {t};
	
\end{tikzpicture}
\caption{Exploiting quantum error correction.
   One share of a $((2,3))$ threshold quantum secret sharing scheme is allocated to each of the call points $y_j$. Meanwhile, each reveal point $z_j$ is lightlike to both $y_j$ and $y_{j+1 \operatorname{mod} 3}$. (The vertical direction is time. The red arrows represent causal, in this case lightlike, curves.) Each solid black line represents a causal diamond. If each share is sent along one arrow, then each diamond will intersect two of the red arrows, meaning each diamond contains sufficient shares to reconstruct the encoded quantum state.
}\label{fig:qec}
\end{figure}

When more spatial dimensions are introduced, delocalization becomes crucial in understanding which configurations of causal diamonds are compatible. 
For a representative example, place three call points at time zero on the vertices of an equilateral triangle with edge lengths $\ell$. Place the reveal points at time $\ell/(2c)$ on the midpoints of the edges, for $c$ the speed of light. Because the call-reveal pairs are lightlike, the diamonds $D_i$ are just line segments, as shown in Figure~\ref{fig:qec}. 
There is no causal curve through the diamonds $D_i$, so the strategy of simply moving the qubit around won't work. Delocalizing the qubit through the use of quantum error correcting codes will, however. It is possible to encode the quantum state $\ket{\varphi}$ into a tripartite Hilbert space $\mathcal{H}_1 \otimes \mathcal{H}_2 \otimes \mathcal{H}_3$ such that the qubit can be recovered even if any one of the corresponding three quantum subsystems is lost. This is known as a $((2,3))$ threshold quantum secret sharing scheme because the quantum information can be recovered from any two of the three subsystems even though no information at all can be recovered from fewer than two~\cite{cleve99}. This encoding is performed at the start point $s$ and then one share is forwarded to each of the call points. For each $j$, if the request is made at $y_j$, then that share is forwarded to $z_j$. Otherwise, that share is forwarded to $z_{j-1 \operatorname{mod} 3}$. By this arrangement the correct reveal point will receive two out the three shares as required to recover the state.

The example reveals one interesting way to delocalize quantum information. To address the full variety of ways in which that information can be replicated in spacetime, the following theorem characterizes every summoning task in Minkowski space as possible or impossible. 

\begin{theorem} \label{thm:summoning}
Summoning is possible if and only if the following conditions hold:
\begin{enumerate}
\item Every reveal point is in the future light cone of the starting point $s$.
\item For each pair $(i,j)$, the diamonds $D_i$ and $D_j$ are causally related, meaning that there exists a causal curve from $D_i$ to $D_j$ or vice versa.
\end{enumerate}
\end{theorem}

In other words, a set of causal diamonds can all contain the same quantum information if and only if the diamonds are all causally related to each other. The two conditions are necessary because they encode the most basic constraints coming from relativity and quantum mechanics, namely causality and the impossibility of cloning. Indeed, Condition 1 is manifestly the prohibition of superluminal communication. Condition 2 arises from reasoning similar to Kent's treatment of ``non-ideal'' summoning~\cite{kent11}.
Suppose there is a successful summoning protocol for which Condition 2 is violated, meaning that two diamonds $D_i$ and $D_j$ are spacelike separated as in Figure 1. If the call is received at $y_i$, there is a procedure that will reveal the state at $z_i$. Now imagine that the call machinery malfunctions such that it makes a call at $y_j$ in addition to the one at $y_i$. Because $y_j$ is not in the causal past of $z_i$, the malfunction cannot prevent the state from being revealed at $z_i$. Likewise, because $y_i$ is not in the causal past of $z_j$, the call at $y_j$ will result in the state successfully being revealed at $z_j$. This procedure therefore reveals the state $\ket{\varphi}$ at the two spacelike points $z_i$ and $z_j$ starting from a single copy of $\ket{\varphi}$ at the point $s$. In other words, a summoning protocol for a configuration violating Condition 2 is easily modified to make a cloning machine, which is impossible.

To see that Conditions 1 and 2 are sufficient will require constructing a protocol that will succeed at the summoning task given a starting point 
and $n$ call-reveal pairs 
satisfying the conditions. The structure of the protocol will only depend on the directed graph $G=(V,E)$ whose vertices are labelled by the diamonds $D_i$ and which contains the  edge $(D_i,D_j)$ if and only if there is a causal curve from some point in $D_i$ to one in $D_j$. 

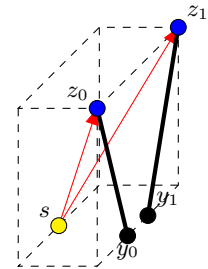
\begin{figure}[t]
\begin{tikzpicture}[scale=0.70]

	\coordinate (PT) at (0,0,0);
	
	\coordinate (CL) at (1.5,0,0.5);
	\coordinate (CR) at (1.5,0,-0.5);
	
	\coordinate (QL) at (1.5,3,2);
	\coordinate (QR) at (1.5,3,-2);
	
	\draw[dashed] (QL) -- (QR) -- (1.5,0,-2) -- (1.5,0,2) -- (QL);
	\draw[dashed] (1.5,0,-2) -- (0,0,-2) -- (0,0,2) -- (1.5,0,2);
	\draw[dashed] (1.5,3,-2) -- (0,3,-2) -- (0,3,2) -- (1.5,3,2);
	\draw[dashed] (0,3,-2) -- (0,0,-2);
	\draw[dashed] (0,3,2) -- (0,0,2);
	
	\draw[red][-triangle 45] (PT) -- (QL);
	\draw[red][-triangle 45] (PT) -- (QR);
	
	\draw[ultra thick] (CL) -- (QL);
	\draw[ultra thick] (CR) -- (QR);
	
	\draw[fill=yellow] (PT) circle (0.15);
	\node [above left] at (PT) {$s$};
	
	\draw[fill=black] (CL) circle (0.15);
	\draw[fill=black] (CR) circle (0.15);
	
	\node [below] at (CL) {$y_0$};
	\node [above right] at (CR) {$y_1$};
	
	\draw[fill=blue]  (QL) circle (0.15);
	\draw[fill=blue]  (QR) circle (0.15);
	
	\node [above left] at (QL) {$z_0$};
	\node [above right] at (QR) {$z_1$};
	
\end{tikzpicture}
\caption{Exploiting teleportation.
  The $n=2$ strategy can be used to complete this example, even though $y_0$ and $y_1$ are outside the light cone of $s$. The essence of teleportation is that it splits a qubit into entanglement and classical data transmission, thereby making it possible to delocalize quantum information in a curious way: classical data can be transmitted to several recipients without regard to the no-cloning theorem while entanglement reaches outside the light cone. Both features are crucial in this example.}\label{fig:teleportation}
\end{figure}

It is possible to handle the $n=2$ case by making use of teleportation~\cite{kent12}. Without loss of generality, assume there is a causal curve from $D_0$ to $D_1$. Begin by distributing a Bell pair between the spatial locations of the start point and $y_0$. Upon receiving the quantum state at the start point, immediately teleport it over the Bell pair~\cite{bennett92}, sending the classical teleportation data to both $z_0$ and $z_1$. Meanwhile, if the call is received at $y_0$, forward the other half of the Bell pair to $z_0$, but if no call is received, forward it to $z_1$. Because there is a causal curve from the start point to both $z_0$ and $z_1$, and because there is a causal curve from $D_0$ and $D_1$ (which, in particular, guarantees there is a causal curve from $y_0$ to $z_1$), both the classical data and the half of the Bell pair required to reconstruct the quantum state will arrive at the appropriate reveal point. Figure~\ref{fig:teleportation} depicts an example in which this protocol succeeds but the simpler strategy of carrying the qubit through the causal diamonds fails. 

Using quantum error correction, a protocol for general $n$ can be built recursively from the protocol for $n=2$. Encode the state $\ket{\varphi}$ at the starting point $s$ in an $((n-1,n))$ threshold secret sharing scheme~\cite{cleve99}. There are $n$ subsets of $\{ D_1, D_2, \ldots, D_n \}$ of size $n-1$.  Assign one of the $n$ shares to each of the subsets and for each subset recursively execute the protocol, now on the smaller subset of size $n-1$. If the request is made at call point $y_j$, then for each of the subsets containing $D_j$, the corresponding protocol will forward its share of the secret to $z_j$. Precisely $n-1$ of the $n$ subsets contain $D_j$, so the state $\ket{\varphi}$ will be recoverable at $z_j$, as required. An example for $n=4$ is sketched in Figures 5a and 5b.
\begin{figure}[t!]
\begin{center}$
\begin{array}{cc}
\begin{tikzpicture}[scale=0.70]
			
	\coordinate (s) at (-1,-3,1.25);
	\node at (s) {a)};
			
	\coordinate (P4) at (1.25,-2.5,1.25);
	
	\coordinate (CA) at (0.5,-0.25,0);
	\coordinate (CB) at (3,-0.25,0);
	\coordinate (CC) at (2.5,0,2.5);
	\coordinate (CD) at (0,0,2.5);
	
	\coordinate (QD) at (0.5,2.25,0); 
 	\coordinate (QB) at (2.75,1.2,1.25);
	\coordinate (QA) at (3,2.25,0);
	\coordinate (QC) at (1.25,1.25,2.5);

	\draw[dashed, thin] (CA) -- (CB) -- (CC) -- (CD) -- cycle;
	\draw[dashed, thin] (0.5,2.25,0) -- (3,2.25,0) -- (2.5,2.5,2.5) -- (0,2.5,2.5) -- cycle;
	
	\draw[dashed, thin] (0,0,2.5) -- (0,2.5,2.5);
	\draw[dashed, thin] (2.5,0,2.5) -- (2.5,2.5,2.5);
	
	\node [below] at (CA) {$y_3$};
	\node [below right] at (CB) {$y_2$};
	\node [below right] at (CC) {$y_1$};
	\node [below left] at (CD) {$y_0$};
	
	\draw[red][-triangle 45] (CC) -- (QB);
	\draw[red][-triangle 45] (CB) -- (QA);
	\draw[red][-triangle 45] (CA) -- (QD);
	\draw[red][-triangle 45] (CD) -- (QC);
	\draw[red][-triangle 45] (CB) -- (QD);
	\draw[red][-triangle 45] (CC) -- (QA);
	
	\draw[very thick] (CA) -- (QA);
	\draw[very thick] (CB) -- (QB);
	\draw[ultra thick] (CC) -- (QC);
	\draw[ultra thick] (CD) -- (QD);
	
	\draw[fill=black] (CA) circle (0.15);
	\draw[fill=black] (CB) circle (0.15);
	\draw[fill=black] (CC) circle (0.15);
	\draw[fill=black] (CD) circle (0.15);
	
	\draw[fill=blue]  (QA) circle (0.15);
	\draw[fill=blue]  (QB) circle (0.15);
	\draw[fill=blue]  (QC) circle (0.15);
	\draw[fill=blue]  (QD) circle (0.15);
	
	\node [above right] at (QA) {$z_3$};
	\node [right] at (QB) {$z_2$};
	\node [above] at (QC) {$z_1$};
	\node [above left] at (QD) {$z_0$};
	
	\draw[fill=yellow] (P4) circle (0.15);
	\node [below right] at (P4) {$s$};
	
\end{tikzpicture} &
\begin{tikzpicture}[scale=0.22]

	\node (t'') at (-7,-10) {}; 
	\node (t') at (-5,-8) {b)};
	\node (t) at (-5,-18) {c)};

	\coordinate (A) at (0,0);
	\coordinate (B) at (7,0);
	\coordinate (C) at (7,-7);
	\coordinate (D) at (0,-7);
	
	\draw (A) circle (0.2);
	\draw (B) circle (0.2);
	\draw (C) circle (0.2);
	\draw (D) circle (0.2);
	
	\draw[-latex] (0.2,0) -> (6.8,0);
	\draw[-latex] (7,-0.2) -> (7,-6.8);
	\draw[-latex] (6.8,-7) -> (0.2,-7);
	\draw[-latex] (0,-6.8) -> (0,-0.2);
	
	\draw[-latex] (6.86,-6.86) -> (0.14,-0.14);
	\draw[-latex] (6.86,-0.14) -> (0.14,-6.86);
	
	\node[left] at (A) {\small{$D_0$}};
	\node[right] at (B) {\small{$D_1$}};
	\node[right] at (C) {\small{$D_2$}};
	\node[left] at (D) {\small{$D_3$}};
	
	\coordinate (P) at (0,-10);
	\coordinate (Q) at (7,-10);
	\coordinate (R) at (7,-17);
	\coordinate (W) at (0,-17);
	
	\coordinate (PQ) at (3.5,-10);
	\coordinate (QR) at (7, -13.5);
	\coordinate (RW) at (3.5, -17);
	\coordinate (WP) at (0, -13.5);
	\coordinate (PR) at (1.75, -11.75);
	\coordinate (WQ) at (5.25, -11.75);
	
	\draw (P) circle (0.2);
	\draw (Q) circle (0.2);
	\draw (R) circle (0.2);
	\draw (W) circle (0.2);
	
	\draw (PQ) circle (0.2);
	\draw (QR) circle (0.2);
	\draw (RW) circle (0.2);
	\draw (WP) circle (0.2);
	\draw (PR) circle (0.2);
	\draw (WQ) circle (0.2);
	
	\draw(0.2,-10) -- (3.3,-10);
	\draw(3.7,-10) -- (6.8,-10);
	
	\draw[dashed] (7,-10.2) -- (7,-13.3);
	\draw[dashed] (7,-13.7) -- (7,-16.8);
	
	\draw[dashed] (6.8,-17) -- (3.7,-17);
	\draw[dashed] (3.3,-17) -- (0.2,-17);
	
	\draw (0,-16.8) -- (0,-13.7);
	\draw (0,-13.3) -- (0,-10.2);
	
	\draw (6.86,-16.86) -- (1.85,-11.89);
	\draw (0.14,-10.14) -- (1.58,-11.58);
	
	\draw[dashed] (6.86,-10.14) -- (5.39,-11.59);
	\draw[dashed] (0.14,-16.86) -- (5.11, -11.88);

	\node[left] at (P) {\small{$D_0$}};
	\node[right] at (Q) {\small{$D_1$}};
	\node[right] at (R) {\small{$D_2$}};
	\node[left] at (W) {\small{$D_3$}};

	\pgfmathsetmacro{\XValueArc}{\ArcRadius*cos(\ArcAngle)}%
	\pgfmathsetmacro{\YValueArc}{\ArcRadius*sin(\ArcAngle)}%

	\pgfmathsetmacro{\XValueLabel}{\ArcRadius*cos(\ArcAngle/2)}%
	\pgfmathsetmacro{\YValueLabel}{\ArcRadius*sin(\ArcAngle/2)}%
	
	\draw[purple] ($(-0.5,-9.5)+(1.7,0)$) 
          arc (0:-90:1.7);
          \draw[purple] ($(-0.5,-17.5)+(1.7,0)$) 
          arc (0:90:1.7);
          \draw[purple] ($(7.5,-9.5) -(1.7,0)$)
          arc (0:90:-1.7);
          \draw[purple] ($(7.5,-17.5) -(1.7,0)$)
          arc (0:-90:-1.7);

\end{tikzpicture}

\end{array}$
\end{center}
\caption{General Strategy.
 a) This configuration of four call-reveal spacetime points satisfies the conditions of Theorem~\ref{thm:summoning} and requires a combination of error correction and teleportation for successful summoning. Each reveal point is again lightlike to its call point and $z_j$ is causal to the call point $y_{j-1 \operatorname{mod} 4}$. In addition, $z_0$ is lightlike to $y_2$ and $z_3$ is lightlike to $y_1$. b) The graph $G$ of causal relationships used to construct the summoning protocol. The subproblem involving the cycle $D_0 \rightarrow D_2 \rightarrow D_3 \rightarrow D_0$ is structurally equivalent to the example illustrated in Figure 3, although the recursive protocol replaces direct quantum communication by teleportation. c) The code used  in the efficient construction has one physical qubit for each edge of this graph, $G'$. The arcs label the four sets of doubled edges sufficient to reconstruct the state. In the protocol, if there is a request at $y_0$, for example, the qubits on the solid edges are sent to $z_0$.}\label{fig:general}
\end{figure}
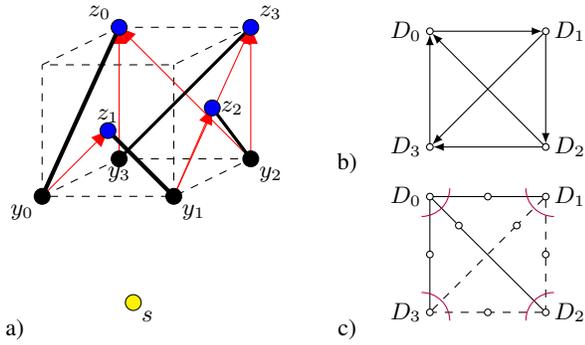

\textbf{Efficient construction}--- The protocol described in the proof of Theorem~\ref{thm:summoning} is unfortunately inefficient, using roughly $n!$ qubits. Practicality aside, such dramatic growth quickly runs afoul of the holographic bound, which places a limit of roughly $1.4 \times 10^{69}$ bits per square meter: trying to store the protocol's qubits in a region centred at $s$ of area of 1 m$^2$ would already create a black hole for $n=55$. (See, e.g., \cite{bousso02}.) Understanding summoning in the presence of gravity therefore requires finding more efficient protocols.

The high cost is incurred from the recursive encoding preceding teleportation, so we will show how to achieve the same functionality directly. To begin, it will be helpful to characterize that functionality. The recursive encoding produces a quantum error correcting code with shares associated to the edges of the graph $G$. Should the call arrive at $y_j$, teleportation is used to ensure that all shares associated to the edges incident to $D_j$ arrive at the reveal point $z_j$. The functionality required of the code is therefore clear: it must be possible to recover the encoded quantum state using only those shares received at $z_j$. That is, the erasure of shares for any combination of edges leaving at least one vertex untouched must be recoverable, as illustrated in Figure 5c. 

That describes an unusual quantum error correcting code. Ignoring directionality, $G$ is the complete graph, so each vertex is incident to exactly $n-1$ edges. The total number of edges is ${n \choose 2}$, however, so the quantum state must be recoverable from a vanishing fraction $2/n$ of the total number of shares, albeit a specially chosen vanishing fraction that prevents violations of the no-cloning principle. A particular codeword-stabilized  (CWS) quantum code~\cite{cross2008codeword} will do the job very efficiently. Let $G'=(V',E')$ be an undirected version of $G$ with a new vertex placed on each edge, so that $V' = V \cup E$ and $E' = \{ ( v, (v,w) ) : (v,w) \in E \text{ or } (w,v) \in E \}$. The code will consist of a single qubit for each of the edges of $G'$. For each such edge $e$, let $N_e = \{ f \in E' : f \cap e \neq \emptyset \mbox{ and } f \neq e \}$ be the set of edges adjacent to $e$ and define $S_e = X_e \prod_{f \in N_e} Z_f$, where $X_e$ and $Z_e$ are Pauli operators acting on edge $e$. 
The encoded qubit is simply the span of the simultaneous $+1$ and simultaneous $-1$ eigenstates of all the $S_e$ operators.

This code has precisely the desired properties if the share for each edge $e \in E$ of the original graph $G$ is identified with the pair of edges replacing $e$ in $G'$ as we demonstrate below. The construction therefore yields a method for solving the summoning task using exactly $n(n-1)$ physical qubits per summoned qubit.

\textbf{Analysis of the CWS code---} The CWS formalism constructs a quantum error correcting code from an undirected graph $H$ and a classical error correcting code, associating a physical qubit with each vertex of $H$~\cite{cross2008codeword}. Since we require a share for each \emph{edge} of $G$, we will choose $H$ to be the line graph of the undirected graph $G'$. The classical error correcting code will simply be a length $2 {n \choose 2}$ binary repetition code. These choices uniquely specify the code described above. Because the binary repetition code forms a group, this CWS code is also a stabilizer code~\cite{cross2008codeword}.

The conditions for quantum error correction in a CWS code can easily be re-expressed in terms of the graph $G'$. CWS codes map all Pauli errors into patterns of $Z$ errors. In particular, the error $X_e$ on any edge $e$ is mapped into the product $\prod_{f \in N_e} Z_f$ of $Z$ errors on all the edges adjacent to $e$. Up to an irrelevant sign, therefore, any Pauli error $P$ induces a product of $Z$ errors $\operatorname{err}(P)$.

As explained earlier, it must be possible to correct against the erasure of any shares leaving any vertex $u$ of $G = (V,E)$ untouched. That is equivalent to being able to correct any set of Pauli errors located on edges of $G'$ not associated with $u$. Since the edges $E'$ of $G'$ are of the form $(v,(v,w))$ for $v \in V$ and $(v,w) \in E$, that amounts to being able to correct errors on any $(v,(v,w))$ such that $u \not\in (v,w)$. Let $\mathcal{E}$ be the set of such errors for any fixed $u$. Since our construction is invariant with respect to permutations of the vertices of $V$, the choice of $u$ doesn't matter.

The error correction conditions specialized to our case~\cite{cross2008codeword} are that, for each $P \in \mathcal{E}$, 
\begin{itemize}
\item $\,$ $\operatorname{err}(P) \neq \prod_e Z_e$ and
\item $\,$ either $\operatorname{err}(P) \neq I$ or $\left[ \, \prod_ e Z_e , P \right] = 0$.
\end{itemize}
The first condition is easy to verify. 
Let $P \in \operatorname{err}(P)$ be a single-qubit error acting on edge $(v,(v,w))$. By assumption, $u \not\in (v,w)$. The only edge adjacent to $(v,(v,w))$ containing $u$ is $(v,(v,u))$. In particular, there are no edges of the form $(u,(u,x))$ adjacent to $(v,(v,w))$ so $P$ cannot induce an error on any such edge, and neither can any product of single-qubit errors in $\mathcal{E}$. Therefore $\operatorname{err}(P) \neq \prod_e Z_e$ for all $P \in \mathcal{E}$.

For the second condition, begin by considering a single-qubit $X$ error $P \in \mathcal{E}$, again acting on edge $(v,(v,w))$. $(v,(v,u))$ is the only edge adjacent to $(v,(v,w))$ containing $u$, so $\operatorname{err}(P)$ will contain exactly one $Z$ operator acting on edges containing $u$. Using that $Z^2 = I$, it follows that any multiqubit $X$ error $P \in \mathcal{E}$ for which $\operatorname{err}(P) = I$ must contain an even number of single-qubit errors. But $XZ = -ZX$ so $\left[ \, \prod_ e Z_e , P \right] = 0$ for any such error $P$. A general multiqubit error will contain arbitrary products of $X$ and $Z$ errors, but since $Z$ errors commute with $\prod_e Z_e$ and don't propagate along edges, the second condition holds for them as well.

\textbf{Discussion}--- 
Our analysis of summoning information provides a detailed characterization of all the ways in which quantum information can be replicated in spacetime. It is remarkable that the sets of causal diamonds that can all hold the same quantum information can be so cleanly characterized:
the conditions of Theorem~\ref{thm:summoning} are simple constraints arising from causality and no-cloning, but they prove to be sufficient. Thus, the most basic restrictions imposed by quantum mechanics and special relativity turn out to be the only restrictions.

These conclusions should extend to more general spacetimes since our arguments have only depended on causal structure. Ultimately, however, the reasoning will likely to break down in a full theory of quantum gravity, for which it has been argued that the interplay between cloning and causality should be much more subtle~\cite{susskind1993stretched,hayden2007black}. Most importantly, the use of no-cloning to rule out summoning for a pair of spacelike separated causal diamonds likely fails in some situations in which the structure of spacetime prevents the two clones from ever being compared.

A question closely related to the problem studied here is whether quantum mechanics and relativity can be combined to build a ``position-based'' cryptography, which was recently answered largely in the negative~\cite{KentM2011,Buhrman2011,buhrman2014position}. In both that work and ours, novel combinations of entanglement and teleportation make it possible to transmit quantum information in ways that initially seem to defy causality. Nonetheless, both scenarios prove to be quite amenable to analysis, and can be seen as building blocks~\cite{Beckman2001,Beckman2002,paw2009information,BeigiK2011,kent12} for a larger theory of distributed computation in relativistic spacetime.

\textit{Note added in proof:} Since this paper first appeared on arXiv.org in 2012, there has been further development of summoning~\cite{adlam2015quantum,hayden2016spacetime} and a surge of exciting work in relativistic quantum information theory more generally. (See e.g. \cite{downes2013quantum,friis2013relativistic,martin2013processing,martin2013sustainable,ahmadi2014quantum,jonsson2015information}.)

\textit{Acknowledgments} We are indebted to Raphael Bousso, Adrian Kent, Fang Xi Lin, Rob Myers and Prakash Panangaden for helpful discussions. This work was supported by the Canada Research Chairs program, FQXi, the Perimeter Institute, CIFAR, NSERC, ONR through grant N000140811249, and the Simons Foundation. The Perimeter Institute is supported by Industry Canada and Ontario's Ministry of Economic Development \& Innovation.

\bibliographystyle{unsrt}
\bibliography{summoning}

\section*{Appendix A - On definitions and assumptions}

There are a number of assumptions implicit in the definition of summoning supplied in the main text. Kent does an excellent job of clearly identifying them in his article, to which we refer the reader for a more thorough discussion~\cite{kent11}. Most importantly, we assume that information at any spacetime point can be transmitted without errors to any other point in its future light cone, and that quantum computations at any point can be performed instantaneously.

Those assumptions are necessary in order to be able to assert that a qubit is localized to the causal diamond $D_j$ if and only if it can be summoned to $z_j$. In a realistic physical theory, features of the dynamics beyond causal structure and quantum mechanics could prevent the summons from succeeding. The quantum state of planet Earth exactly one day ago, for example, is localized within a spatial sphere one light-day in radius, but it is doubtful that the state could really be reassembled here on Earth one day hence.

Summoning with a separation between the call and reveal point as we've defined it in this article is actually a version of  ``non-ideal summoning'' in Kent's terminology, which he analyzed for a case in which all the call points are on one spacelike hyperplane and all the reveal points on another. Kent introduced the separation in order to allow time to process information between the receipt of the request and the need to reveal the state. Because our emphasis has been on the ways quantum information can be localized in principle, our interpretation of the separations has been slightly different, but nothing prevents using them for Kent's original purpose.

A final point requiring comment is that localizing even a single qubit to a point or even a causal diamond with lightlike separated call and reveal points would violate the holographic bound~\cite{bousso02}. For the purposes of this article, therefore, a point should just be regarded as a volume of space small with respect to the other length scales being discussed.

\end{document}